\documentclass[preprint,pra,nofootinbib]{revtex4}
\usepackage{mathrsfs}
\usepackage{graphicx}
\usepackage{epsfig}
\usepackage{dcolumn}
\usepackage{bm}
\usepackage{amsmath,amssymb,amsthm}
\usepackage[colorlinks=true,linkcolor=blue]{hyperref}
\usepackage{subfigure}
\usepackage{booktabs}
\usepackage[mathscr]{euscript}
\usepackage[normalem]{ulem}
\UseRawInputEncoding

\newcommand\redsout{\bgroup\markoverwith{\textcolor{red}{\rule[0.6ex]{6pt}{0.6pt}}}\ULon}

\begin{document}


\title{Ren-integrable and ren-symmetric integrable systems}
\author{S. Y. Lou
}
\affiliation{
\mbox {\scriptsize School of Physical Science and Technology, Ningbo University, Ningbo, 315211, China }}



\date{\today}

\begin{abstract}
A new type of symmetry, ren-symmetry describing anyon physics and the corresponding topological physics, is proposed.
		Ren-symmetry is a generalization of super-symmetry which is widely applied in super-symmetric physics such as the super-symmetric quantum mechanics, super-symmetric gravity, super-symmetric string theory, super-symmetric integrable systems and so on.
		The super-symmetry and Grassmann-number are, in some sense, the dual conceptions,
		which turns out that these conceptions coincide for the ren situation, that is, a similar conception of ren-number is devised to ren-symmetry.
		In particular, some basic results of the ren-number and ren-symmetry are exposed which allow one to derive, in principle, some new types of integrable systems including ren-integrable models and ren-symmetric integrable systems.
		Training examples of ren-integrable KdV type systems and ren-symmetric KdV equations are explicitly given.	
\end{abstract}


\maketitle
\section{Introduction}

The idea of symmetry originates in natural scientific fields, and its importance there is well known.
Symmetry considerations belong to the most universal and astonishing methods by which scientists have successfully solved the problems in building new solutions from known ones \cite{Sol}, doing dimensional reductions of nonlinear partial differential equations \cite{Sol1,Sol2,Sol3,Sol4}, getting new integrable systems \cite{Sol5,Sol6,Sol7,Sol8} and even constructing all solutions for certain nonlinear systems \cite{Sol9}.

By using the SU(3)$\times$SU(2)$\times$U(1) symmetry, three fundamental interactions, the strong, weak and electromagnetic interactions, have been unified into the so-called standard model. However, in order to unify the gravitational interaction to the standard model, one has to introduce a new type of symmetry, say the super-symmetry between bosons and fermions.
New areas of physical fields including the super-symmetric gravity \cite{SG,SG1}, super-symmetric quantum mechanics \cite{SM}, super-symmetric string theory \cite{SS} and super-symmetric integrable systems \cite{SusyI,SusyII,SusyIII,SusyIV,SusyV,SusyVI} have been developed which are highly motivated by super-symmetries,  in the belief that they possess a high potential for future development.

In super-symmetry theory, it is essential to introduce the Grassmann variable $\theta$ \cite{GA,GA1} and the super-symmetric derivative ${\cal{D}}$ with the properties
\begin{eqnarray}
&& \theta^2=0,\ \theta_1\theta_2=-\theta_2\theta_1,\label{GA}\\
&& {\cal{D}}= \partial_{\theta}+\theta\partial_x,\ {\cal{D}}^2=\partial_x.\label{calD}
\end{eqnarray}
The super-symmetric derivative ${\cal{D}}$ is invariant under the super-symmetric transformation
\begin{eqnarray}
&&\theta\rightarrow \theta+\eta,\ x\rightarrow x-\theta \eta.
\label{STrans}
\end{eqnarray}

Recently, different to the bosons and fermions, anyons with fractional charges, spin and statistics in two dimensions have been attracted high sufficient attentions by many scientists \cite{any1,any2,any3,any4}. Anyons can be used to describe some kinds of quasi-particles (the low-energy excitations in Hamiltonian
systems) including the fractional quantum Hall states \cite{qHall}, vortices
in topological superconductors \cite{topo} and Majorana zero modes in semiconductors proximitized by superconductors \cite{Majo}.
By analogy with fermion case in which fermions can be described by Grassmann fields, some new fields endowed to describe anyons,  we call anyon-fields and/or ren-fields. We shall use the adjective ``ren" to stress the arbitrary of $\alpha$ and to avoid the confusion on ``arbitrary symmetry" or ``any symmetry". ``Ren" means ``arbitrary" in Chinese.

A comparison of the Grassmann number $\theta=\sqrt{0}$ and the super-symmetric derivative ${\cal{D}}=\sqrt{\partial_x}$, corresponding to ¡°ren¡± point of view, suggests one to use, as the ren-number and the ren-symmetric derivative, respectively, the following radical generalization of the formulae
$$\theta_{\alpha}\equiv\theta=\sqrt[\alpha]{0},~~ {\cal{R}}=\sqrt[\alpha]{\partial_x}$$
with $\alpha$ being arbitrary.

 The introduction of the G-number and the super-symmetric derivative yields some significant novel mathematical and physical fields such as the Grassmann algebra \cite{GA}, the super-symmetric quantum mechanics \cite{SM}, the super-symmetric string theory \cite{SS}, the super-symmetric gravity \cite{SG}, the super/Kuper-integrable systems \cite{SI,SI1,SI2,SI3} and super-symmetric integrable theories \cite{SusyI}.
Therefore, we hope that the introduction of the ren-number and the ren-symmetric derivative, may successfully create some different mathematical and physical fields such as the ren-algebra, ren-calculus, ren integrable models and ren-symmetric integrable systems. The usual G-number, G-algebra, super-symmetric theory, super-integrable systems and super-symmetric integrable systems just correspond to the ren-case for $\alpha=2$.

In Sec. II of this paper, the concept of the R-number $\theta$ for an arbitrary positive integer $\alpha$ is defined with the aim of deriving ren-algebra, ren-derivative, and ren-symmetric derivative.
Then, we deal in Sec. III with the problem of finding some types of ren-integrable systems by coupling the usual boson fields and the anyon fields.
When $\alpha$ is fixed to $\alpha=2$, the ren-integrable system is just the known super- or Kuper-integrable systems \cite{SI}. By means of the ren-symmetric derivative, we study the ren-symmetric integrable systems in Sec. IV, the ren-symmetric KdV systems for $\alpha=3,\ 4$ are explicitly given. The well known super-symmetric integrable systems are just the special cases of the ren-symmetric integrable systems with $\alpha=2$. The last section includes a short summary and some discussions.

\section{Ren-algebra, ren-derivative and ren-symmetric derivative}
\bf Definition 1. \rm A ren-number $\theta\equiv \theta_{\alpha}$ is defined as a number possessing the properties
\begin{equation}
\theta^{\alpha}=0,\ \theta^i\neq 0,\ i=1,\ 2,\ \ldots,\ \alpha-1, \label{theta}
\end{equation}
where $\alpha$ is an arbitrary positive integer.

This is, a ren-number can be spelled out as a non-zero $\alpha$ root of zero
\begin{equation}
\theta=\sqrt[\alpha]{0}\neq0. \label{theta0}
\end{equation}
It is clear that there are $\alpha-1$ solutions of \eqref{theta0}, $\{\theta,\ \theta^2,\ \ldots,\ \theta^{\alpha-1}\}$.
Such a definition, exists in one important special case $\alpha=2$, the usual G-number $\theta=\theta_2$.

For $\alpha=2$, we know that if $a$ and $b$ are Grassmann numbers then the combination is still a G-number when the anti-communication relation
\begin{equation}
ab=-ba \label{anti}
\end{equation}
holds.

Similarly, for $\alpha\geq 2$, if $a$ and $b$ are ren-numbers with the $q$-commutation relation ($\mbox{\rm i}\equiv\sqrt{-1}$),
\begin{eqnarray}\label{q}
&&ab=q_jba,\quad q_j^{\alpha}=1,\ q_j=q^j,\ q=\exp\left(\frac{2 \pi \mbox{\rm i}}{\alpha}\right), \nonumber\\
&&\qquad j\in {\cal{J}_{\alpha}}= \{1,\ 2,\ \ldots,\ \alpha-1\}/\{n_mp_m<\alpha\},
\end{eqnarray}
then so is the combination $a+b$ of $a$ and $b$,
where $\{n_mp_m<\alpha\}$ is a set with $p_m,\ m=1,\ \ldots,\ M$ being the prime factors of $\alpha$ and $n_m=1,\ \ldots,\ N_m$ being integers with $N_mp_m<\alpha$. For $\alpha=3,\ 4,\ \ldots,\ 10$, we have
\begin{eqnarray}\label{J}
&& {\cal{J}}_{3}= \{1,\ 2\},\  {\cal{J}}_{4}= \{1,\ 3\},\  {\cal{J}}_{5}= \{1,\ 2,\ 3,\ 4\},\  {\cal{J}}_{6}= \{1,\ 5\},\ {\cal{J}}_{7}= \{1,\ 2,\ \ldots,\ 6\},\  \nonumber\\
&& {\cal{J}}_{8}= \{1,\ 3,\ 5,\ 7\},\  {\cal{J}}_{9}= \{1,\ 2,\ 4,\ 5,\ 7,\ 8\},\  {\cal{J}}_{10}= \{1,\ 3,\ 7,\ 9\}.
\end{eqnarray}

From the expression of $q$ given in \eqref{q}, we know that the usual number (boson number) is related to $\alpha=\infty$ and the Grassmann number (fermion number) corresponds to $\alpha=2$.

\bf Definition 2. \rm The degree, $\beta \ (\mbox{\rm mod}(\alpha))$, of a ren-number $\gamma$ is defined as
\begin{equation}\label{order}
\gamma\theta=q^{\beta}\theta\gamma, \\
\end{equation}
where the degree of $\theta$ is always fixed as one in this paper.
$\gamma$ with \eqref{order} is said also a $\beta$ order ren-number.

If the ren-numbers $\gamma_1$ and $\gamma_2$ possess the degrees $\beta_1$ and $\beta_2$, respectively, then we have the commutation relation
\begin{equation}\label{order12}
\gamma_1\gamma_2=q^{\beta_1\beta_2}\gamma_2\gamma_1, \\
\end{equation}
which is consistent with \eqref{order} when $\beta_2=1$ and $\beta_1\beta_2=0$.

\bf Definition 3. \rm Ren-derivative, $\frac{\mbox{\rm d}}{\mbox{\rm d} \theta} $, is a derivative with respect to the ren-variable $\theta$,
\begin{equation}
\frac{\mbox{\rm d} f(\theta)}{\mbox{\rm d} \theta}=\left.\frac{f(\theta_1)-f(\theta)}{\theta_1-\theta}\right|_{\theta_1\rightarrow \theta}=\frac{f(\theta)-f(q\theta)}{(1-q)\theta}. \label{ftheta}
\end{equation}

Similar to the Grassmann case, because the definition of the ren-number \eqref{theta}, an arbitrary function of ren-variable, $f(\theta)$ can be written as
\begin{equation}
f(\theta)=\sum_{i=0}^{\alpha-1}\theta^i g_i=\sum_{i=0}^{\alpha-1}f_i\theta^i. \label{fexp}
\end{equation}
If $f(\theta)$ is a $\beta$ order ren-number, then $f_i$ and $g_i$ in \eqref{fexp} are $\beta-i$ order ren-numbers with $f_i=q^{i\beta-i^2}g_i$.

According to the property \eqref{fexp}, it is enough to find all the possible ren-derivatives for an arbitrary ren-function $f(\theta)$ by calculating
$$\frac{\mbox{\rm d} \theta^i}{\mbox{\rm d} \theta},\ i=1,\ 2,\ \ldots,\ \alpha-1.$$

Based on the commutation relation \eqref{q} and the definition of the ren-derivative \eqref{ftheta}, it is readily to prove that
\begin{equation}
\frac{\mbox{\rm d} \theta^i}{\mbox{\rm d} \theta}=\sum_{k=0}^{i-1}q^k\theta^{i-1}=\frac{1-q^i}{1-q}\theta^{i-1}\equiv i_q\theta^{i-1}, \label{dtheta}
\end{equation}
where $i_q$ is defined as $i_q=1+q+\cdots+q^{i-1}$, say, $2_q=1+q,\ 3_q=1+q+q^2$ and so on.

Thus, for the ren-function $f(\theta)$ with degree  $\beta$, we have
\begin{equation}
\frac{\mbox{\rm d}f(\theta)}{\mbox{\rm d}\theta } =\sum_{i=0}^{\alpha-1} i_q\theta^{i-1}g_i =\sum_{i=0}^{\alpha-1}\frac{(q^{\beta-i}-q^{\beta})}{1-q}f_i\theta^{i-1}
=\sum_{i=0}^{\alpha-1}q^{\beta-i}i_qf_i\theta^{i-1}. \label{df}
\end{equation}
The ren-integration may be defined as an inverse operator of the ren-derivative for $\theta^k,\ k<\alpha-1$, however, for $\theta^{\alpha-1}$ the inverse operator of the ren-derivative is not well defined. A different integration operator can be defined \cite{Majid}. For $\alpha=2$, the Berezin integral has been defined \cite{Berezin,Bin}. In this paper, we will not discuss this problem though the similar Berezin integral may be introduced under the requirement of the translation invariance \cite{Majid}.

\bf Definition 4. \rm A ren-symmetric derivative $\cal{R}\equiv \cal{R}_{\alpha}$ is defined as an $\alpha$ root of the usual space derivative $\partial_x$, i.e.,
\begin{equation}
{\cal{R}^{\alpha}}=\partial_x,\ {\cal{R}}=\sqrt[\alpha]{\partial_x}. \label{Rdef}
\end{equation}
It is interesting that in terms of the ren-number $\theta$, the ren-symmetric derivative $\cal{R}$ can be explicitly written as
\begin{equation}
{\cal{R}}=\partial_{\theta}+\frac1{[(\alpha-1)!]_q}\theta^{\alpha-1}\partial_x, \label{Rres}
\end{equation}
where $[n!]_q$ is defined as
\begin{eqnarray}
[n!]_q&\equiv&\prod_{i=1}^{n}\frac{1-q^{i}}{1-q}\equiv \prod_{i=1}^{n} i_q
\equiv 1_q\cdot2_q\cdots (n-1)_q\cdot n_q, \label{nq!}
\end{eqnarray}
whose particular case $q=1$ is the usual $n!$.

It is reasonable that the ren-symmetric derivative \eqref{Rres} will reduce back to the known super-symmetric derivative ${\cal R}_2\equiv {\cal D} =\partial_{\theta}+\theta\partial_x $ when $\alpha=2$. The ren-symmetric derivatives for $\alpha=3,\ 4,\ 5,\ 6$ and $7$ are given by the formulae as a straightforward computation,
\begin{eqnarray}
&&{\cal R}_3=\partial_{\theta}+\frac1{[2!]_q}\theta^2\partial_x=\partial_{\theta}-q\theta^2\partial_x,\ q=\mbox{\rm e}^{\left({2\pi \mbox{\rm \tiny i}}/3 \right)}=\frac12(\sqrt{3}\mbox{\rm i}-1),\nonumber\\
&&{\cal R}_4=\partial_{\theta}+\frac{1}{[3!]_q}\theta^3\partial_x=\partial_{\theta} -\frac12(1+q)\theta^3\partial_x,\ q=\mbox{\rm e}^{\left({2\pi \mbox{\rm \tiny i}}/4 \right)}=\mbox{\rm i},\nonumber\\
&&{\cal R}_5=\partial_{\theta}+\frac{1}{[4!]_q}\theta^4\partial_x=\partial_{\theta}+\frac{q^3}{(1+q)^2}\theta^4\partial_x,
\ q=\mbox{\rm e}^{\left( {2\pi \mbox{\rm \tiny i}}/5 \right)},\nonumber\\
&&{\cal R}_6=\partial_{\theta}+\frac{1}{[5!]_q}\theta^5\partial_x =\partial_{\theta}+\frac{q}6\theta^5\partial_x,\  q=\mbox{\rm e}^{\left({2\pi \mbox{\rm \tiny i}}/6 \right)}=\frac12(1+\sqrt{3}\mbox{\rm i}),\nonumber\\
&&{\cal R}_7=\partial_{\theta}+\frac{1}{[6!]_q}\theta^6\partial_x =\partial_{\theta}-\frac{q^6}{[3!]_q^2}\theta^6\partial_x,\  q=\mbox{\rm e}^{\left({2\pi \mbox{\rm \tiny i}}/7 \right)}.
\label{R34}
\end{eqnarray}

It is not difficult to prove that the ren-symmetric derivative ${\cal{R}}$ possesses the following ren-symmetric transformation
\begin{equation}
\theta\rightarrow \theta+\eta,\ x\rightarrow x-f(\theta,\ \eta), \label{rensymm}
\end{equation}
with
\begin{equation}
f=f(\theta,\ \eta)= \sum_{k=1}^{\alpha-1}\frac{1}{[k!]_q[(\alpha-k)!]_q}\theta^{k}\eta^{\alpha-k}. \label{renf}
\end{equation}
With the stress on the first few $f$, say, $\alpha=2,\ 3,\ 4,\ 5$ and $6$, we have
\begin{eqnarray}
&f=\theta \eta,& \alpha=2,\nonumber\\
&f=\frac1{[2!]_q}(\theta\eta^2+\theta^2\eta), & \alpha=3,\nonumber\\
&f=\frac1{[3!]_q}\left(\theta \eta^3+\frac{3_q}{2_q}\theta^2\eta^2
+\theta^3\eta\right), & \alpha=4,\nonumber\\
&f=\frac1{[4!]_q}\left(\theta\eta^4+\frac{4_q}{2_q}\theta^2\eta^3
+\frac{4_q}{2_q}\theta^3\eta^2+\theta^4\eta\right), & \alpha=5,\nonumber\\
&f=\frac1{[5!]_q}\left(\theta\eta^5+\frac{5_q}{2_q}\theta^2\eta^4
+\frac{5_q4_q}{3_q2_q}\theta^3\eta^3+\frac{5_q}{2_q}\theta^4\eta^2+\theta^5\eta\right), &\alpha=6. \label{f26}
\end{eqnarray}

\section{Ren-integrable systems}
In the limit $\alpha=2$, ren-integrable models are just the known super- or Kuper- integrable models which are first proposed by Kupershmidt in \cite{SI}.
It is known that the usual bosonic KdV equation,
\begin{equation}
u_t=(-u_{xx}+3u^2)_x, \label{KdV}
\end{equation}
	is determined as a compatibility condition $[L,\ S]=LS-SL=0$ of equations from a Lax pair
\begin{eqnarray}
&&\psi_{xx}-(u+\lambda)\psi\equiv L\psi=0,\nonumber\\
&&\psi_t-3u_x\psi-6u\psi_x+4\psi_{xxx}\equiv S\psi=0. \label{KdVLax}
\end{eqnarray}

Usually, the spectral function $\psi$ is considered as a boson function. In fact, because the Lax pair \eqref{KdVLax} is linear, the spectral function $\psi$ may be a fermion function and even a ren function.

It is known that if $\sigma$ is a symmetry of an integrable evolution equation
\begin{equation}
u_t=K(u), \label{Ku}
\end{equation}
i.e., a solution of
\begin{equation}
\sigma_t=K'\sigma \equiv \lim_{\epsilon=0} \partial_{\epsilon}K(u+\epsilon \sigma), \label{Kusym}
\end{equation}
then
\begin{equation}
u_t=K(u)+\sigma, \label{Kus}
\end{equation}
is also an integrable model.

Furthermore, if $\sigma=\sigma(\psi)$, where $\psi$ is a spectral function of the Lax pair,
\begin{equation}
\hat{L}\psi=0,\qquad \hat{S}\psi=0, \label{LSKu}
\end{equation}
of \eqref{Ku}, then
the first type of source equation
\begin{eqnarray}
&&u_t=K(u)+\sigma(\psi),\nonumber\\
&&\hat{L}\psi=0, \label{Kupsi1}
\end{eqnarray}
and the second type of source equation
\begin{eqnarray}
&&u_t=K(u)+\sigma(\psi),\nonumber\\
&&\hat{S}\psi=0, \label{Kupsi2}
\end{eqnarray}
may all be integrable \cite{Source1,Source2,Source3,Source4,Source5,Source6,Source7,Source8,Source9}.

Usually, the spectral functions $\psi$ studied in the integrable models are restricted as bosonic functions. For instance, for the KdV equation \eqref{KdV} the first and second types of integrable bosonic source equations possess the forms
\begin{eqnarray}
&&u_t=(-u_{xx}+3u^2+\langle\phi|\phi\rangle)_x,\qquad \langle\phi|\phi\rangle\equiv \sum_{i=1}^n \phi_i^2,\nonumber\\
&&(\partial_x^2-u-\lambda_i)\phi_i=0, \quad i=1,\ 2,\ \ldots,\ n,  \label{KdVphi1}
\end{eqnarray}
and
\begin{eqnarray}
&&u_t=(-u_{xx}+3u^2+\langle\phi|\phi\rangle)_x,\nonumber\\
&&(\partial_t+4\partial_x^3-6u\partial_x-3u_x)\phi_i=0, \quad i=1,\ 2,\ \ldots,\ n,  \label{KdVphi2}
\end{eqnarray}
respectively.

Now, if we extend the spectral function of the Lax pair \eqref{KdVLax} to a ren-function, $\xi$, then we have a trivial symmetry, $\sigma=\xi^{\alpha-1}\xi_{xx}$. Applying this symmetry to the second type of source equation, we can find some coupled ren systems
\begin{eqnarray}
&&u_t=(-u_{xx}+3u^2)_x+\sum_{i=1}^k \langle\xi_i^{\alpha_i-1}|\xi_{i,xx}\rangle,\qquad \langle\xi_i^{\alpha_i-1}|\xi_{i,xx}\rangle\equiv \sum_{j=1}^{n_i} \xi_{ij}^{\alpha_i-1}\xi_{ij,xx},\nonumber\\
&&(\partial_t+4\partial_x^3-6u\partial_x-3u_x)\xi_{ij}=0, \quad j=1,\ 2,\ \ldots,\ n_i,\ i=1,\ 2,\ \ldots,
\ k,\  \xi_{ij}^{\alpha_{i}}=0   \label{KdVxi}
\end{eqnarray}
with one boson field $u$ and $k\times n_i$ ren fields $\xi_{ij}$, where $k,\ n_i$ and $\alpha_i$ are all arbitrary integers.

The integrability of \eqref{KdVxi} with $\alpha_i=2$ for all $i=1,\ 2,\ \ldots,\ k$ is known because the models reduce back to the so-called super/Kuper-integrable systems \cite{SI,SI1,SI2,SI3}. Before studying the integrability of \eqref{KdVxi} with $\alpha_i\neq 2$ for some $2 <i\leq k$, we directly write down a more general nontrivial symmetry of the KdV equation \eqref{KdV},
\begin{eqnarray}
&&\sigma=(\xi_1\xi_{2x}-\xi_{1x}\xi_2)_x,   \label{sxi12}
\end{eqnarray}
where $\xi_1$ and $\xi_2$ are ren-spectral functions of the usual KdV equation with the same spectral parameter $\lambda=\lambda_1=\lambda_2$ but with different degrees, $\beta$ and $\alpha-\beta$, respectively.

The simplest second type of source equation related to \eqref{sxi12}  possesses the form
\begin{eqnarray}
&&u_t=[3u^2-u_{xx}+12(\xi_1\xi_{2x}-\xi_{1x}\xi_2)]_x,
\nonumber\\
&&(\partial_t+4\partial_x^3-6u\partial_x-3u_x)\xi_{i}=0,\ i=1,\ 2. \label{RKdV}
\end{eqnarray}
\bf Theorem. \rm  The model \eqref{RKdV} is Lax integrable with the Lax pair
\begin{eqnarray}
&&\psi_{xx}-(u+\lambda)\psi+\left(\int \xi_1\psi \mbox{\rm d}x\right) \xi_2
-\xi_1 \left(\int \xi_2\psi \mbox{\rm d}x\right) \equiv \hat{L}\psi =0,
\nonumber\\
&&(\partial_t+4\partial_x^3-6u\partial_x-3u_x)\psi\equiv \hat{S}\psi=0. \label{RLax}
\end{eqnarray}
\bf Proof. \rm To complete the proof of the theorem, it suffices to prove that the compatibility condition
\begin{eqnarray}
[\hat{L},\ \hat{S}]f=(\hat{L}\hat{S}-\hat{S}\hat{L})f \label{LSf}
\end{eqnarray}
is valid for arbitrary $f$ if \eqref{RKdV} is satisfied.

Expanding the expression \eqref{LSf} with the operators $\hat{L}$ and $\hat{S}$ defined in \eqref{RLax},
\begin{eqnarray}
&&\left(\int\xi_1f\mbox{\rm d}x\right) (6u\xi_{2x}+3\xi_2u_x-4\xi_{2xxx}-\xi_{2t})+(\xi_{1t}+4\xi_{1xxx}-6u\xi_{1x}-3\xi_1u_x)
\int\xi_2f\mbox{\rm d}x\nonumber\\
&&
\qquad +\xi_1\left[\int (\xi_{2t}f-4 \xi_2f_{xxx} +6 \xi_2uf_x+3\xi_2u_xf)\mbox{\rm d} x -8\xi_{2xx}f-4\xi_{2x}f_x\right]\nonumber\\
&&
\qquad -\left[\int (\xi_{1t}f-4\xi_1f_{xxx}+6\xi_1uf_x+3\xi_1u_xf)\mbox{\rm d} x -8\xi_{1xx}f-4\xi_{1x}f_x\right] \xi_2\nonumber\\
&&\qquad +(u_t-6uu_x+u_{xxx})f=0,
 \label{LSf1}
\end{eqnarray}
and simplifying the result with the formulae of integration by parts
\begin{eqnarray}
&&\int\xi_i f_{xxx}\mbox{\rm d} x = \xi_i f_{xx}-\xi_{ix}f_x+\xi_{ixx}f-\int\xi_{ixxx}f\mbox{\rm d}x, \nonumber\\
&&\int\xi_i u f_x \mbox{\rm d} x = \xi_i u f-\int f (u\xi_i)_x\mbox{\rm d}x,
 \label{parts}
\end{eqnarray}
\eqref{LSf1} is changed to
\begin{eqnarray}
&&\int\xi_1f\mbox{\rm d}x\   (6u\xi_{2x}+3\xi_2u_x-4\xi_{2xxx}-\xi_{2t})+(\xi_{1t}+4\xi_{1xxx}-6u\xi_{1x}-3\xi_1u_x)
\int\xi_2f\mbox{\rm d}x\nonumber\\
&&
+\xi_1\int (\xi_{2t}+4 \xi_{2xxx}-6 u \xi_{2x}- 3\xi_2u_x) f\mbox{\rm d} x
-\left[\int (\xi_{1t}+4\xi_{1xxx}-6u\xi_1-3\xi_1u_x)f\mbox{\rm d} x\right]\xi_2  \nonumber\\
&& +(u_t-6uu_x+u_{xxx}+12\xi_{1xx}\xi_2-12\xi_1\xi_{2xx})f=0.
 \label{LSf2}
\end{eqnarray}
Because of the arbitrariness of $f$, \eqref{LSf2} is correct only after joining to it the equation \eqref{RKdV}.
The theorem is proved.

\bf Remark. \rm From the proof procedure of the theorem, it is known that we have not used any commutation relation on $\xi_1$ and $\xi_2$. That means \eqref{RKdV} is Lax integrable no matter the fields $\xi_1$ and $\xi_2$ are boson fields, fermion fields and/or ren-fields with arbitrary $\alpha$.
\section{Ren-symmetric integrable systems}
In Sec. II of this paper, we have defined the ren-symmetric derivative $\cal{R}$. By means of the ren-symmetric derivative, the usual bosonic integrable systems can be extended to ren-symmetric integrable ones. Before discussing ren-symmetric integrable systems, we list some special cases for $\alpha=2$, i.e., the super-symmetric integrable models.
\subsection{Super-symmetric integrable KdV systems}
The most general $N=1$ symmetric form of the KdV equation \eqref{KdV} is generated by the fermionic super-field $\Phi$ with an arbitrary constant $a$,
\begin{equation}
\Phi_t+\Phi_{xxx}+a({\cal{D}}\Phi_x)\Phi+(6-a)({\cal{D}}\Phi)\Phi_x=0. \label{susyKdV}
\end{equation}
 Mathieu had proven that the super-symmetric KdV equation \eqref{susyKdV} is Painlev\'e integrable only for $a=0$ and $3$ \cite{Mathieu}. Although the super-symmetric KdV system \eqref{susyKdV} is not Painlev\'e integrable for arbitrary $a$, it does possess multiple soliton solutions \cite{KdVa}.  In \eqref{susyKdV}, the super-field $\Phi\equiv \xi+\theta u $ is a fermionic super-field with a fermion field $\xi$ and a boson field $u$.

For the coupled KdV equation, we have an interacting model between a susy-boson field $U$ and a susy-fermion field $\Phi$
\begin{eqnarray}
&& \Phi_t=(-\Phi_{xx}+3\Phi {\cal D}\Phi+6U\Phi)_x,\nonumber\\
&& U_t=(-U_{xx}+3U^2+3\Phi{\cal D}U )_x, \label{PhiU}
\end{eqnarray}
which is Lax integrable. Incorporating $U=0$, \eqref{PhiU} readily reduces back to \eqref{susyKdV}  with $a=3$.
For $\Phi=0$, \eqref{PhiU} becomes a quite trivial extension of the original KdV equation \eqref{KdV} by $u\rightarrow U$.

The component form of \eqref{PhiU} reads
\begin{eqnarray}
&& v_t=(-v_{xx}+3v^2+3\zeta_x\zeta+6uv+6\xi\zeta)_x,\nonumber\\
&& u_t=(-u_{xx}+3u^2+3\zeta\xi)_x,\nonumber\\
&& \xi_t=(-\xi_{xx}+6u\xi+3v\xi+3u_x\zeta)_x,\nonumber\\
&& \zeta_t=(-\zeta_{xx}+6u\zeta+3v\zeta)_x, \label{Phiu}
\end{eqnarray}
with $U=u+\theta \xi$ and $\Phi=\zeta+\theta v$, where $u$ and $v$ are boson components and $\xi$ and $\zeta$ are fermion components.

The Lax pair of \eqref{PhiU} can be written as
\begin{eqnarray}
&& \Psi_{xx}=\Phi {\cal D}\Psi+(U+\lambda)\Psi,\nonumber\\
&& \Psi_t=-4\Psi_{xxx}+6U\Psi_x+3U_x\Psi+6\Phi{\cal D}\Psi_x+3\Phi_x{\cal D}\Psi. \label{PhiLax}
\end{eqnarray}

\subsection{Ren-symmetric integrable KdV systems}
Analogous to \eqref{susyKdV}, the general ren-symmetric KdV equation is expressible in the form
\begin{equation}\label{renKdV}
\Phi_t+\Phi_{xxx}+\sum_{i=0}^{\left[\beta_1\right]}a_i\left({\cal{R}}^i\Phi\right)
\left({\cal{R}}^{\alpha+\beta-i}\Phi\right)=0, \ \beta=0,\ 1,\ 2,\ \ldots,\ \alpha-1,\ \beta_1\equiv \frac{\alpha+\beta}2,
\end{equation}
where $\left[\beta_1\right]$ is the integer part of $\beta_1$, $a_i,\ i=0,\ 1,\ 2,\ \ldots,\ [\beta_1]$, are arbitrary bosonic constants, $\beta$ is the degree of the ren-field $\Phi\equiv \Phi(x,\ t,\ \theta)$.

As in the super-symmetric ($\alpha=2$) case, one may find some possible integrable cases by fixing the constants $a_i$ of the ren-symmetric KdV equation \eqref{renKdV}.  For instance, the Lax integrable systems,
\begin{equation}
\Phi_{jt}+\Phi_{jxxx}-3{\cal{R}}^{\alpha-j}\left({\cal{R}}^{j}\Phi_j\right)^2+\rho_j=0,\ {\cal{R}}^{j}\rho_j=0, \ j=0,\ 1,\ \ldots,\ \alpha-1,\ \label{phij}
\end{equation}
with $\rho_j=0$ possessing Lax pair of
\begin{eqnarray}\label{renLax}
&&\Psi_{xx}-({\cal{R}}^{j}\Phi_j+\lambda)\Psi=0,\nonumber\\
&&\Psi_{t}+4\Psi_{xxx}-3\left({\cal{R}}^{j}\Phi_{jx}\right)\Psi-6\left({\cal{R}}^{j}\Phi_j\right)\Psi_x=0,
\end{eqnarray}
 are just the special cases of \eqref{renKdV}. The degrees of $\Phi_j$ and $\rho_j$ in \eqref{phij} are $j$.

For $\alpha=3$,
the ren-symmetric KdV equation \eqref{renKdV} becomes
\begin{eqnarray}
&&\Phi_{0t}+\Phi_{0xxx}+a\Phi_0\Phi_{0x}+b({\cal{R}}\Phi_0)({\cal{R}}^2\Phi_0)=0,\ {\cal{R}}=\partial_{\theta}-q\theta^2\partial_x, \label{b0}\\
&&\Phi_{1t}+\Phi_{1xxx}+a\Phi_1({\cal{R}}\Phi_{1x})+b({\cal{R}}\Phi_1)\Phi_{1x}
+c({\cal{R}}^2\Phi_1)^2=0,\label{b1}\\
&&\Phi_{2t}+\Phi_{2xxx}+a\Phi_2({\cal{R}}^2\Phi_{2x})+b({\cal{R}}\Phi_2)({\cal{R}}\Phi_{2x})
+c({\cal{R}}^2\Phi_2)(\Phi_{2x})=0,\label{b2}
\end{eqnarray}
where $a,\ b$ and $c$ are arbitrary constants and $\Phi_0,\ \Phi_1$ and $\Phi_2$ are the ren-fields with degrees, $0,\ 1$ and $2$, respectively.

The special integrable case \eqref{phij} for $\{\alpha=3, j=0\}$ is related to \eqref{b0} with $b=0$ up to a re-scaling. \eqref{b1} with $\{a=0,\ c=b\}$ is equivalent to the integrable case \eqref{phij} for $\{\alpha=3, j=1\}$. Taking $\{a=b=0\}$ in \eqref{b2} leads to the equivalent special integrable ren-symmetric KdV equation \eqref{phij} with $\{\alpha=3, j=2\}$.

Incorporating the explicit forms for $$\Phi_0=u+\theta  \zeta+\theta^2 \xi,\ \Phi_1=\xi+\theta u+\theta^2 \zeta,\ \Phi_2=\zeta+\theta \xi+\theta^2 u$$
	and the consistent commutation relations
	\begin{equation}\label{xz}
	\xi\theta=q\theta\xi,\ \zeta\theta=q^2\theta\zeta,\ \xi\zeta=q^2\zeta\xi,\ \zeta\zeta_x=q\zeta_x\zeta,\ \zeta\zeta_{xx}=q\zeta_{xx}\zeta
	\end{equation} leads to the coupled component forms of \eqref{b0},\ \eqref{b1} and \eqref{b2}
\begin{eqnarray}
&&u_t+u_{xxx}+auu_x-q^2b\zeta\cdot \xi=0, \nonumber\\
&&\zeta_t+\zeta_{xxx}+a(\zeta u)_x+b\zeta u_x+bq\xi^2=0,\nonumber\\
&&\xi_t+\xi_{xxx}+a(u\xi)_x+(a-b)\zeta_x\cdot \zeta=0,\label{b01}
\end{eqnarray}
\begin{eqnarray}
&&\xi_t+\xi_{xxx}+a\xi u_x+bu\xi_x+cq\zeta^2=0,\nonumber\\
&&u_t+u_{xxx}+(a+b)uu_x-aq^2\zeta_x\cdot \xi-(2cq^2+b) \xi_x\cdot \zeta=0,\nonumber\\
&&\zeta_t+\zeta_{xxx}+(a-bq-cq^2)\zeta u_x+(b-aq^2)u\zeta_x-aq\xi_{xx}\cdot \xi+q(c-b)\xi_x^2=0,\label{b11}
\end{eqnarray}
and
\begin{eqnarray}
&&\zeta_t+\zeta_{xxx}-aq^2\zeta u_x+b\xi\cdot \xi_x-cq^2u\zeta_x=0, \nonumber\\
&&u_t+u_{xxx}-q(aq+cq-b)uu_x+(aq-b)\xi\cdot \zeta_{xx}-(bq^2+cq-c) \xi_{x}\cdot \zeta_x-aq\xi_{xx}\cdot \zeta=0,\nonumber\\
&&\xi_t+\xi_{xxx}-(aq^2+b)\xi u_x-(b+c)q^2u\xi_x+a\zeta_{xx}\cdot \zeta+c\zeta_x^2=0,\label{b21}
\end{eqnarray}
respectively. $u$ in \eqref{b01}--\eqref{b21} is a bosonic field and $\xi$ and $\zeta$ are ren-fields with degrees $1$ and $2$, respectively.

The known special integrable ren-symmetric KdV systems of \eqref{b01} and \eqref{b11} read
\begin{eqnarray}
&&u_t+u_{xxx}-6uu_x=0,\nonumber\\
&&\xi_t+\xi_{xxx}-6(u\xi)_x-6\zeta_x\cdot \zeta=0,\nonumber\\
&&\zeta_t+\zeta_{xxx}-6(\zeta u)_x=0, \label{b02}
\end{eqnarray}
and
\begin{eqnarray}
&&u_t+u_{xxx}-6uu_x-6q(1-q) \xi_x\cdot \zeta=0,\nonumber\\
&&\xi_t+\xi_{xxx}-6u\xi_x-6q\zeta^2=0,\nonumber\\
&&\zeta_t+\zeta_{xxx}-6(\zeta u)_x=0,\label{b12}
\end{eqnarray}
respectively.
The special integrable ren-symmetric KdV system of \eqref{b21} is equivalent to that of \eqref{b11} by the transformation $\zeta_x\rightarrow \zeta$.

The choice $\alpha=4,\ q=\mbox{\rm i},\ {\cal{R}}=\partial_{\theta}-\frac{1+q}2\theta^3\partial_x$ leads ren-symmetric KdV equation \eqref{renKdV} straightforwardly to
\begin{eqnarray}
&&\Phi_{0t}+\Phi_{0xxx}+a\Phi_0\Phi_{0x}+b({\cal{R}}\Phi_0)({\cal{R}}^3\Phi_0)
+c({\cal{R}}^2\Phi_0)^2=0,\label{B0}\\
&&\Phi_{1t}+\Phi_{1xxx}+a\Phi_1{\cal{R}}\Phi_{1x}+b({\cal{R}}\Phi_1)\Phi_{1x}
+c({\cal{R}}^2\Phi_1)({\cal{R}}^3\Phi_1)=0,\label{B1}\\
&&\Phi_{2t}+\Phi_{2xxx}+a\Phi_2{\cal{R}}^2\Phi_{2x}+b({\cal{R}}\Phi_2)({\cal{R}}\Phi_{2x})
+c({\cal{R}}^2\Phi_2)(\Phi_{2x})+d({\cal{R}}^3\Phi_2)^2=0,\label{B2}\\
&&\Phi_{3t}+\Phi_{3xxx}+a\Phi_3{\cal{R}}^3\Phi_{3x}+b({\cal{R}}\Phi_3)({\cal{R}}^2\Phi_{3x})
+c({\cal{R}}^2\Phi_3)({\cal{R}}\Phi_{3x})+d\Phi_{3x}{\cal{R}}^3\Phi_3=0,\label{B3}
\end{eqnarray}
where $\Phi_0,\ \Phi_1,\ \Phi_2$ and $\Phi_3$ are the ren-fields with the degrees, $0,\ 1,\ 2$ and $3$, respectively.

\section{Summary and discussions}

In retrospect, the usual Grassmann number and the super-symmetric derivative have been straightforwardly extended to more general forms, the ren-number and the ren-symmetric derivatives, to be applicable to describe physically important quasi-particles, anyons.
Applying the ren-numbers and ren-symmetric derivatives to integrable theory, we have extended the super-integrable and super-symmetric integrable systems to ren-integrable and ren-symmetric integrable systems.

 It is interesting that the ren-integrable KdV system \eqref{RKdV} possesses the completely same form for arbitrary $\alpha$ even for the boson case ($\alpha=\infty$) and fermion case ($\alpha=2$). The only difference is that the degrees of the ren-fields $\xi_1$ and $\xi_2$ should be complementary, say, $\beta$ and $\alpha-\beta$, such that $\xi_1\xi_2$ becomes a boson.

The ren-integrable system \eqref{RKdV} can be further extended to
\begin{eqnarray}
&&u_t=\left[3u^2-u_{xx}+\langle\phi|\phi\rangle
+12\sum_{\alpha=2}^\infty\sum_{\beta_{\alpha}=0}^{\alpha-1}
(\langle\xi_{\{\alpha,\beta_\alpha\}}|\zeta_{\{\alpha,\alpha-\beta_{\alpha}\},x}\rangle
-\langle\xi_{\{\alpha,\beta_\alpha\},x}|\zeta_{\{\alpha,\alpha-\beta_{\alpha}\}}\rangle)\right]_x,
\nonumber\\
&&(\partial_t+4\partial_x^3-6u\partial_x-3u_x)|\phi\rangle=0,\nonumber\\
&&(\partial_t+4\partial_x^3-6u\partial_x-3u_x)|\xi_{\{\alpha,\beta_\alpha\}}\rangle=0,\nonumber\\
&& (\partial_t+4\partial_x^3-6u\partial_x-3u_x)|\zeta_{\{\alpha,\alpha-\beta_{\alpha}\}}\rangle=0,\label{RKdVE}
\end{eqnarray}
where $|\phi\rangle$ is a boson vector field, $|\xi_{\{\alpha,\beta_\alpha\}}\rangle$ is a $\beta_{\alpha}$ order ren-vector field and $|\zeta_{\{\alpha,\alpha-\beta_{\alpha}\}}\rangle$ is an $\alpha-\beta_{\alpha}$ order ren-vector field. The general ren-integrable KdV type system \eqref{RKdVE} is still a Lax integrable model.

Although,  indeed the number of papers produced so far on construction of solutions is incredibly large,  it is necessary to develop some novel methods, one of which may be the so-called bosonization method \cite{GA1}, to construct special solutions of the ren-integrable KdV system \eqref{RKdV} (or more generally \eqref{RKdVE}) and the ren-symmetric KdV system \eqref{renKdV}.

Ren-numbers may also be used to find other types of integrable models such as the dark equations and integrable couplings \cite{darkP,dP}. The concept of dark equations is first introduced by Kupershmidt in Ref. \cite{Dark} where many types of dark KdV systems are given. The modified dark KdV equations are studied in \textcolor{blue}{Ref.} \cite{dark1}. The bosonization procedure of the super-symmetric systems have offered some new types of dark integrable systems \cite{GA1}. The bosonization of ren-symmetric integrable models may yield further dark integrable equations. In fact, applying the bosonization assumptions
\begin{eqnarray}
\xi=\eta p,\ \zeta=\eta^2 q \label{solution}
\end{eqnarray}
with the $\{x,\ t\}$-independent ren-number $\eta$ and the $\{x,\ t\}$-dependent boson fields $p$ and $q$ on the integrable systems \eqref{b02} and/ot \eqref{b12} yields a same standard dark equation system,
\begin{eqnarray}
&&u_t+u_{xxx}-6uu_x=0,\nonumber\\
&&p_t+p_{xxx}-6(up)_x=0,\nonumber\\
&&q_t+q_{xxx}-6(uq)_x=0, \label{darkKdV}
\end{eqnarray}
because of $\eta^3=0$. From \eqref{darkKdV}, we know that the ren-integrable systems \eqref{b02} and \eqref{b12} possess a special types of exact solutions with $u$ being an arbitrary solutions of the usual KdV equation and $\xi$ and $\zeta$ being given by \eqref{solution} while $p$ and $q$ are arbitrary symmetries of the usual KdV equation.

The dark systems can also be considered as some special type of integrable couplings \cite{Ma,Ma1}.
The more about the ren-integrable, ren-symmetric integrable and dark integrable systems should be further studied later.

\begin{acknowledgments}
The work was sponsored by the National Natural Science Foundations of China (Nos.12235007, 11975131). The author is indebt to thank Profs. Q. P. Liu, B. F. Feng, X. B. Hu, R. X. Yao, M. Jia and Drs. K. Tian, X. Z. Hao and D. D. Zhang for their helpful discussions.
\end{acknowledgments}


\end{document}